\documentclass{elsarticle}

\usepackage{hyperref}
\usepackage{subfigure}

\let\today\relax
\makeatletter
\def\ps@pprintTitle{%
    \let\@oddhead\@empty
    \let\@evenhead\@empty
    \def\@oddfoot{\footnotesize\itshape
         {Submitted preprint} \hfill\today}%
    \let\@evenfoot\@oddfoot
    }
\makeatother

\begin{document}

\begin{frontmatter}

\title{Controlling Movement at Nanoscale: Curvature Driven Mechanotaxis}%

\author[UFRN]{Leonardo D. Machado}

\author[USP]{Rafael A. Bizao}

\author[UTrento,QMary]{Nicola M. Pugno\corref{cor2}}
\ead{nicola.pugno@unitn.it}

\author[Unicamp]{Douglas S. Galv{\~a}o \corref{cor1}}
\ead{galvao@ifi.unicamp.br}

\cortext[cor2]{Corresponding author}
\cortext[cor1]{Corresponding author}

\address[UFRN]{Departamento de F\'isica Te\'orica e Experimental, Universidade Federal do Rio Grande do Norte, Natal-RN, 59072-970, Brazil.}

\address[USP]{Institute of Mathematics and Computer Sciences, University of S{\~a}o Paulo, S{\~a}o Carlos, S{\~a}o Paulo, Brazil}

\address[UTrento]{Laboratory of Bio-inspired, Bionic, Nano, Meta Materials \& Mechanics, Department of Civil, Environmental and Mechanical Engineering, University of Trento, Trento, Italy.}

\address[QMary]{School of Engineering and Materials Science, Queen Mary University of London, London, United Kingdom.}

\address[Unicamp]{Instituto de F\'isica "Gleb Wataghin", Universidade Estadual de Campinas, C. P. 6165, 13083-970 Campinas SP, Brazil.}

\begin{abstract}

Locating and manipulating nano-sized objects to drive motion is a time and effort consuming task. Recent advances show that it is possible to generate motion without direct intervention, by embedding the source of motion in the system configuration. In this work, we demonstrate an alternative manner to controllably displace nano-objects without external manipulation, by employing spiral-shaped carbon nanotube (CNT) and graphene nanoribbon structures (GNR). The spiral shape contains smooth gradients of curvature, which lead to smooth gradients of bending energy. We show these gradients can drive nanoscillators. We also carry out an energy analysis by approximating the carbon nanotube to a thin rod and discuss how torsional gradients can be used to drive motion. For the nanoribbons, we also analyzed the role of layer orientation. Our results show that motion is not sustainable for commensurate orientations, in which AB stacking occurs. For incommensurate orientations, friction almost vanishes, and in this instance, the motion can continue even if the driving forces are not very high. This suggests that mild curvature gradients, which can already be found in existing nanostructures, could provide mechanical stimuli to direct motion.
 
\end{abstract}

\end{frontmatter}

\section{Introduction}

Controlling object motion at the nanoscale has proven to be difficult. For instance, although movement can be driven by nanomanipulators \cite{regan2004carbon,zhao2010mass} or Scanning Tunneling/Atomic Force Microscope tips \cite{junno1995controlled, kudernac2011electrically}, locating and displacing nano-sized objects demands time and exquisite control. Other proposals include applying strain gradients \cite{wang2014motion,chen2018nanoscale}, and displacing graphene membranes out-of-plane to induce motion in-plane \cite{dai2016nanoscale}. But manipulation is also required to create these gradients. Alternatives exist that preclude precise manipulation, such as the application of periodic optical or acoustic fields to guide chiral nanoparticles \cite{nourhani2015guiding}. Recent molecular dynamics (MD) simulations \cite{chang2015nanoscale} revealed another option that does not require direct control. In this approach, graphene nanoribbon (GNR) movement is driven by a substrate stiffness gradient. This process is the nanoscale analogue of the biological \textit{durotaxis} - cell migration towards stiffer substrate regions - and was named \textit{nanodurotaxis}. A key advantage of this method is that the motion source is embedded into the system setup. Because of this, no external driving agent is required, the configuration itself provides the energy for movement. \cite{bigoni2015eshelby,bigoni2014torsional}

In this work, based on MD simulations, we demonstrate an alternative manner to controllably displace nano-objects without external manipulation, by creating Archimedean-like spiral shaped
carbon nanotubes (CNT) and GNR. The spiral shape creates smooth curvature gradients, which leads to smooth bending energy gradients and sustained oscillatory movements.

\section{Methodology}
\label{methodology}

Our fully atomistic MD simulations were carried out using the well-known CHARMM \cite{mackerell1998all} force field. The CHARMM potential is a reasonable choice since we are considering large structures ($ \sim 100000$ carbon atoms) and running long MD simulations ($ \sim 5$ nanoseconds). Also, it is important to stress that in our structures, even at their most curved regions, no bond ever reached 5\% strain. Therefore, we remained within the elastic regime for both materials \cite{peng2008measurements,zhao2009size}, where the CHARMM approximations are valid. We have previously used a similar methodology, and succeed in reproducing experimental results \cite{machado2013dynamics,shadmi2015defect}. The integration of the equations of motion was carried out within the NVE ensemble using the LAMMPS MD package \cite{plimpton1995fast}. Our systems were non-periodic and a time-step of 1 femtosecond was used in all simulations.

\section{Results and Discussion}

\begin{figure}
\centering
\includegraphics[width= \columnwidth]{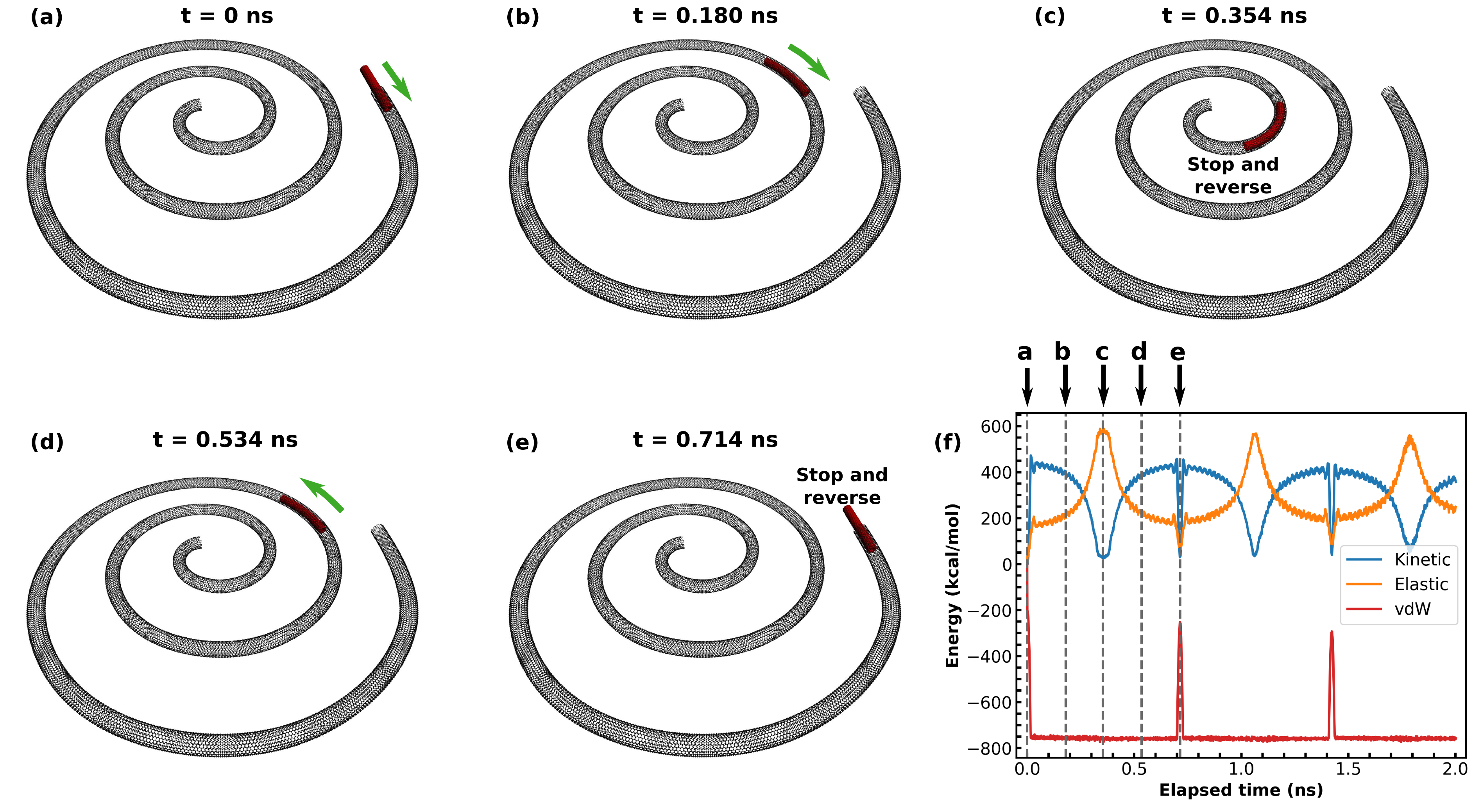}
\caption{(Color online) Molecular dynamics snapshots from a simulation of a CNT nanoscillator. (a) At t = 0 ns, the vdW potential energy is high because the contact area between the inner and outer nanotubes is low. (b) As the CNT is drawn inside, this potential energy is converted to kinetic energy, and motion is driven towards the spiral center. This leads to increased elastic energy. (c) At t = 0.354 ns, the motion is reversed, and (d) conversion of elastic potential energy into kinetic energy starts. (e) At t = 0.714 ns, one oscillation cycle is complete. See also movie S1 \cite{SM}. (f) Evolution of the kinetic, elastic and vdW energy values during the operation of this CNT oscillator. VdW and elastic energy values refer to differences from the initial values. The arrows indicate the energy values associated with the five snapshots presented here. See also movie S1 \cite{SM}. }
\label{fig1}
\end{figure}

The set up used for the nanotube simulations consisted of an inner 8 nm (9,0) CNT inside an outer 200 nm (10,10) CNT, that is frozen into an Archimedes' spiral shape.  The initial simulation configuration consists of the inner tube partially inserted into the outer one (see figure \ref{fig1}) with no provided kinetic energy. The potential energy is high in this configuration, since the contact area ($A$) is small, and the attractive interaction between the tubes is proportional to this area and the adhesion energy ($\Gamma$) \cite{koenig2011ultrastrong}: $E_{ad}=-\Gamma \times A$. Due to van der Waals forces, the inner tube is spontaneously pulled into the outer one.  As the inner CNT moves inwards, it increases its contact area, decreases its potential energy, and increases its kinetic energy. After the inner CNT is fully enveloped by the external CNT, moving further towards the spiral center increases its elastic energy. This happens because the curvature is progressively steeper in an Archimedes' spiral as one approaches the center. The connection between curvature ($\kappa$) and elastic energy can be better understood with the aid of the classical theory of elasticity: for a thin elastic rod with Young's modulus $Y$ and area moment of inertia $I$, the bending energy per unit length is given by $\epsilon_b= YI \kappa^2 /2$ \cite{jawed2014coiling}. As the curvature and the elastic energy increase, the kinetic energy decreases. Eventually, the kinetic energy reaches zero and the inner CNT stops. The movement is then reversed, and the tube moves towards regions of lower elastic energy. As long as the energy dissipation is small, a sustained oscillatory regime is possible \cite{legoas2003molecular}.
Typical simulation results are shown in figure \ref{fig1} and in movie S1 \cite{SM}.

Quantitative results for the energy changes during operation of the nanoscillator described above are presented in figure \ref{fig1} (f). The conversion of van der Waals (vdW) potential energy to kinetic energy occurs in a few picoseconds. For the second period of oscillation, for example, the inner CNT goes partially out and then returns in less than 30 ps. This fast acting restoring force has been shown from MD simulations to be able to drive nanoscillators at frequencies as high as 38 GHz \cite{legoas2003molecular}. In comparison, the conversion of kinetic energy to elastic potential energy is a much slower process in our simulations, and the oscillatory frequency is 1.4 GHz. Part of the total energy is transferred to the internal degrees of atomic motion during oscillations, as evidenced by the decrease in size of the vdW energy peaks from one cycle to the next. 

\begin{figure}
\centering
\includegraphics[width= \columnwidth]{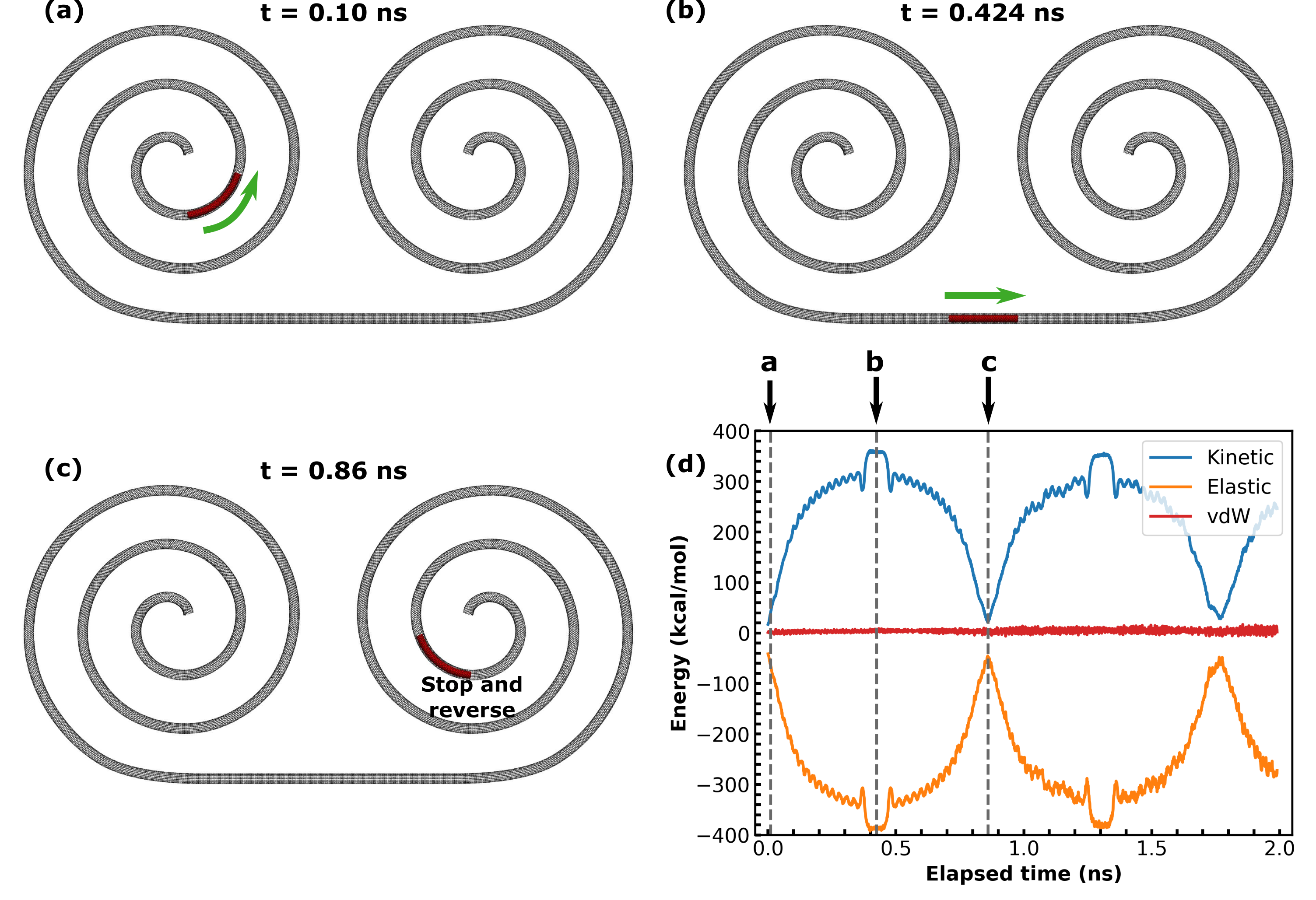}
\caption{(Color online) (a) Snapshots from a molecular dynamics simulation of a CNT nanoscillator. (a) At t $= 0.10$ ns, the elastic potential energy is high because the local Archimedes' spiral curvature is high. (b) As the simulation progresses, the CNT moves towards the right spiral. (c) As the nanotube moves towards the center of right spiral, its kinetic energy decreases, and it eventually stops at t $= 0.86$ ns. Motion is then reversed. (d) Evolution of the kinetic, elastic and vdW energies during the operation of this CNT oscillator. Notice the vdW potential energy remains constant in this setup, because the contact area is constant. The arrows indicate the energy values associated with the five snapshots presented here. See also movie S2 \cite{SM}.}
\label{fig2}
\end{figure}

It is also possible to create nanoscillators that are driven purely by elastic potential energy. One possible configuration to accomplish this is shown in figure \ref{fig2}, consisting of two interconnected CNT spirals. In this setup, the movable nanotube starts in a region of high elastic energy within the left spiral, and as the simulation proceeds this energy is progressively converted into kinetic energy. Motion is at first directed towards regions of lower curvature, and then towards the interconnecting region. As the inner nanotube approaches the center of the right spiral, elastic energy increases, halting the motion. The movement direction is then inverted and, as energy dissipation is small, oscillatory motion is possible. This process is more easily visualized with the aid of movie S2 \cite{SM}). The decrease in the elastic energy peaks between cycles is not as obvious in this case, suggesting a smaller rate of energy transfer to the inner degrees of motion. This oscillator presents a frequency of 1.1 GHz.

Although our discussion has so far focused on energetics, a force-based analysis can provide a qualitative framework relating our work, the \textit{nanodurotaxis}, and previous instances of nanoscillators. If we once again think of CNTs as thin elastic rods, then we should include a torsional component in the total elastic energy: $\epsilon_t=GI\tau^2 /2$. In this equation, $\epsilon_t$ is the twist energy per unit length, $G$ is the shear modulus, $I$ is the area moment of inertia, and $\tau$ is the twist \cite{jawed2014coiling}. Let us also define an attractive energy per unit length: $\epsilon_{ad}= -\gamma A$, where $\gamma=\Gamma / L$. Then the potential energy of the nanotube is

\begin{equation}
    \epsilon_{pot} = \epsilon_{b} +\epsilon_{t}+\epsilon_{ad}. 
    \label{eqenergies}
\end{equation}

Let us also suppose that our nanoscale system as a whole is only free to move along some coordinate $q$. The resulting force on our object is
$$ F = -\frac{d \epsilon_{pot}}{dq} = \gamma \frac{dA}{dq} + A \frac{d \gamma}{dq} - YI \kappa \frac{d \kappa}{dq} - GI\tau \frac{d \tau}{dq}. $$ Let us analyze each of the four force terms that appear in the equation above. The first term corresponds to the case studied by Legoas \textit{et al.} \cite{legoas2003molecular}, in which motion is driven by an increasing contact area. The second term corresponds to the case studied by Chang \textit{et al.} \cite{chang2015nanoscale}: it was shown in this paper that vdW attraction is larger for stiffer substrates, and motion in this instance was driven by a gradient of attractive potential. The third term corresponds to the results presented in this paper, in which motion is driven towards regions of lower curvature. The fourth term corresponds to motion driven towards regions of lower twist, which has yet to be demonstrated at the nanoscale. Even at the macroscale demonstrations of self-propulsion driven by torsion are rather recent \cite{bigoni2014torsional}.

The above analysis suggests other manners in which self-propulsive motion can be achieved. For instance, there might be manners other than gradients of substrate stiffness to create gradients of attractive energy. Likewise, a spiral shape was used in the present paper because it conveniently provided a smooth gradient of curvature. However, any configuration that provides a gradient of bending (or torsional) energy can be used to drive motion. Note also that the different forms of propulsion above can be mixed and matched. Finally, the above framework would lead us to expect that our proposed oscillation mechanism could work for any elastic material. To demonstrate the last point, we designed graphene nanoscillators driven by gradients of bending energy. These are discussed below. Before that, let us verify whether we can attribute the reversal of the nanotube movement when it is inside the spiral entirely to the increased bending energy.

\begin{figure}
\centering
\includegraphics[width= \columnwidth]{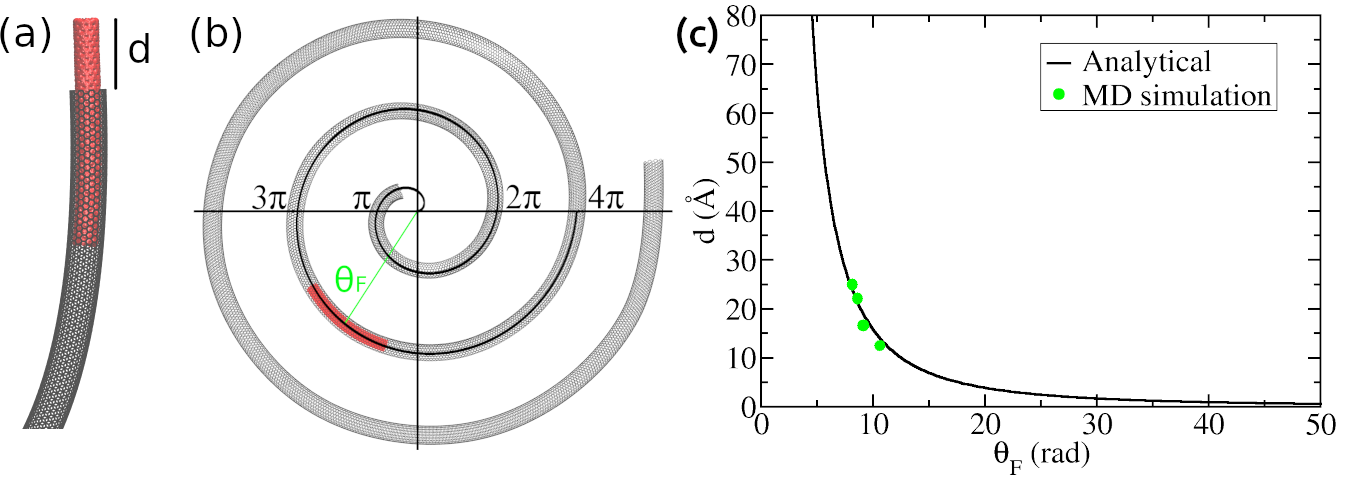}
\caption{(Color online) Snapshots (a) and (b) display the 
initial ejection length $d$ and the final angular position $\theta_F$. Note that for an Archimedean spiral the angle $\theta$ is higher for segments far from the origin, and that $\theta$ increases by 2$ \pi$ whenever the spiral crosses the positive x axis. (c) The solid line corresponds to an analytical curve obtained using Eq. \ref{eqan}. To obtain the green dots, we started MD simulations with different $d$ values and then used the MD trajectories to measure the final $\theta_F$ values. Each simulation corresponds to a green dot. }
\label{figana}
\end{figure}

If the above hypothesis is true, then the initial adhesion energy should have been entirely converted into elastic bending energy at the moment the CNT stops inside the spiral. Consider the inner CNT moving from the initial position showed in Figure \ref{figana}a (with initial ejection length d) to the reversal position showed in Figure \ref{figana}b (where the angle is $\theta_F$). Since the kinetic energy is zero at the two instants examined, Eq. \ref{eqenergies} only has $\epsilon_{ad}$ and $\epsilon_b$ (disregarding small torsional effects). Let us define L as the length of the inner nanotube and let us consider the nanotube curvature ($\kappa$) constant along its length (this approximation is valid for short nanotubes). Then, the initial adhesion energy should be equal to the final bending energy,
$$ (1/2)YI\kappa^2L = \Gamma2\pi rd.$$ Isolating $\kappa$, we find
\begin{equation}
    \kappa = \left(\frac{\Gamma4\pi rd}{YIL}\right)^{1/2}.
    \label{eqan}
\end{equation}
Note that the curvature of the Archimedes' Spiral depends on the $\theta$ angle as
\begin{equation}
    \kappa = \frac{2 + \theta^2}{a(1+\theta^2)^{3/2}}.
    \label{k_theta}
\end{equation}
In this way, we can find the dependence between the initial ejection length d and the final angle $\theta_F$ using Eqs. \ref{eqan} and \ref{k_theta},
\begin{equation}
    \left(\frac{\Gamma4\pi rd}{YIL}\right)^{1/2} = \frac{2 + \theta_F^2}{a(1+\theta_F^2)^{3/2}}.
\end{equation}
With the above equation it is possible to plot $d$ against $\theta_F$, and this curve is displayed using a solid line in Figure \ref{figana}(c). Details on the values used for $\Gamma$, $r$, $Y$, $I$, and $L$ are given in the supporting information \cite{SM}. To verify whether the above expression is valid, we executed four MD simulations, varying the initial ejection length ($d$) and determining the final angle ($\theta_F$). The values obtained are displayed as green dots in Figure \ref{figana}(c). Observe the good agreement between theory and simulation, indicating that motion is indeed driven by bending energy.

\begin{figure}
\centering
\includegraphics[width= \columnwidth]{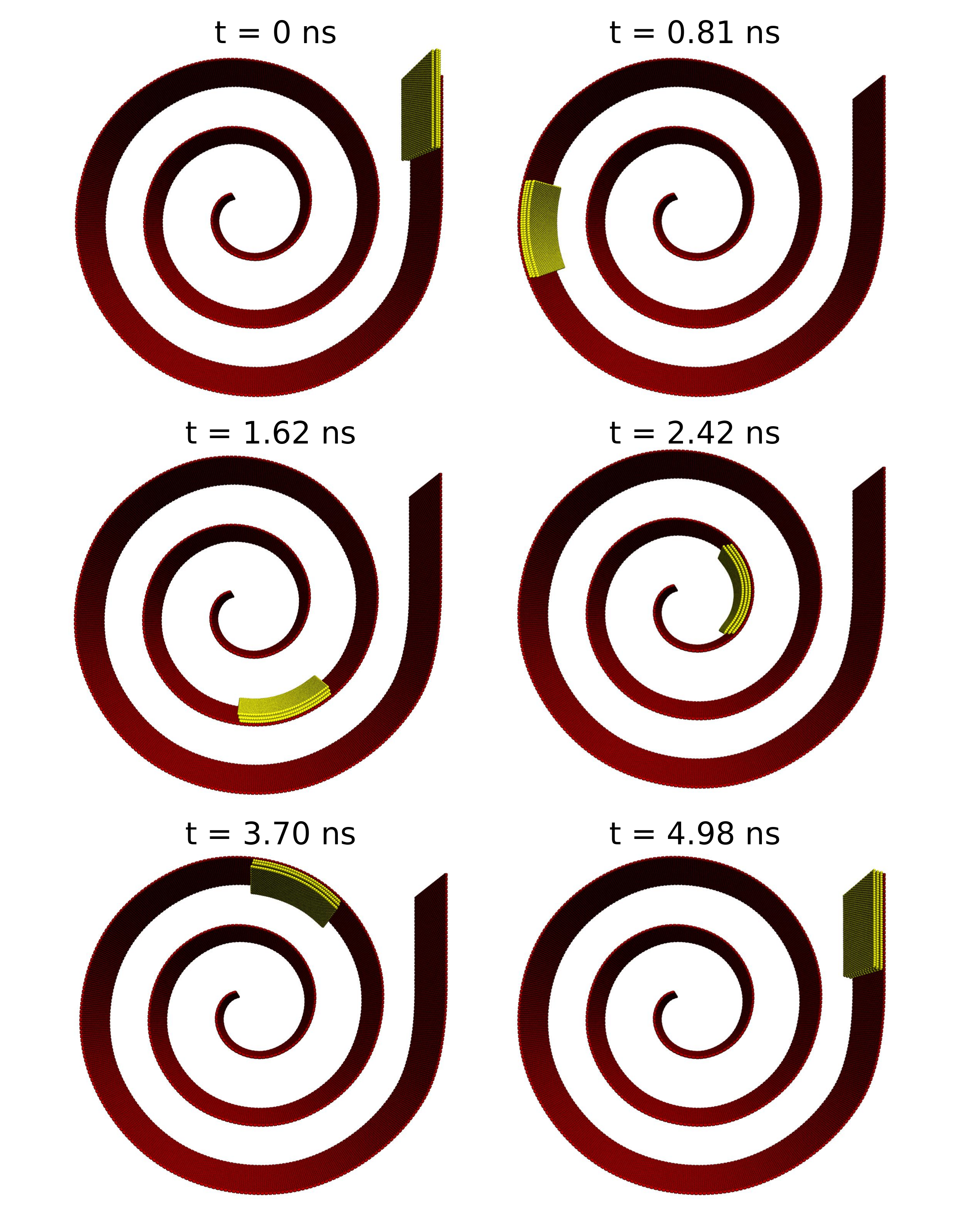}
\caption{(Color online) Snapshots from a MD simulation of a GNR nanoscillator operation. All three GNRs that make up the moving element are zigzag, while the supporting substrate is armchair. Conversion of energy during the operation of CNT and GNR oscillators follows the same sequence. It takes, however, seven times longer to complete one cycle for the latter. See also movie S3 \cite{SM}.}
\label{fig3}
\end{figure}

Snapshots of a MD simulation of a nanoribbon oscillator driven by contact area/curvature gradients is presented in figure \ref{fig3}. In all graphene simulations a 200 nm long 5 nm wide armchair GNR was kept fixed, and was used as substrate for the moving ribbon. We have tested: (i) armchair and zigzag configurations; (ii) single layer (SL), double layer (DL), and triple layer (TL) configurations. We typically found frequencies of $\sim 200$ MHz for our GNR nanoscillators.

\begin{figure}
\centering
\includegraphics[width= 2.5 in]{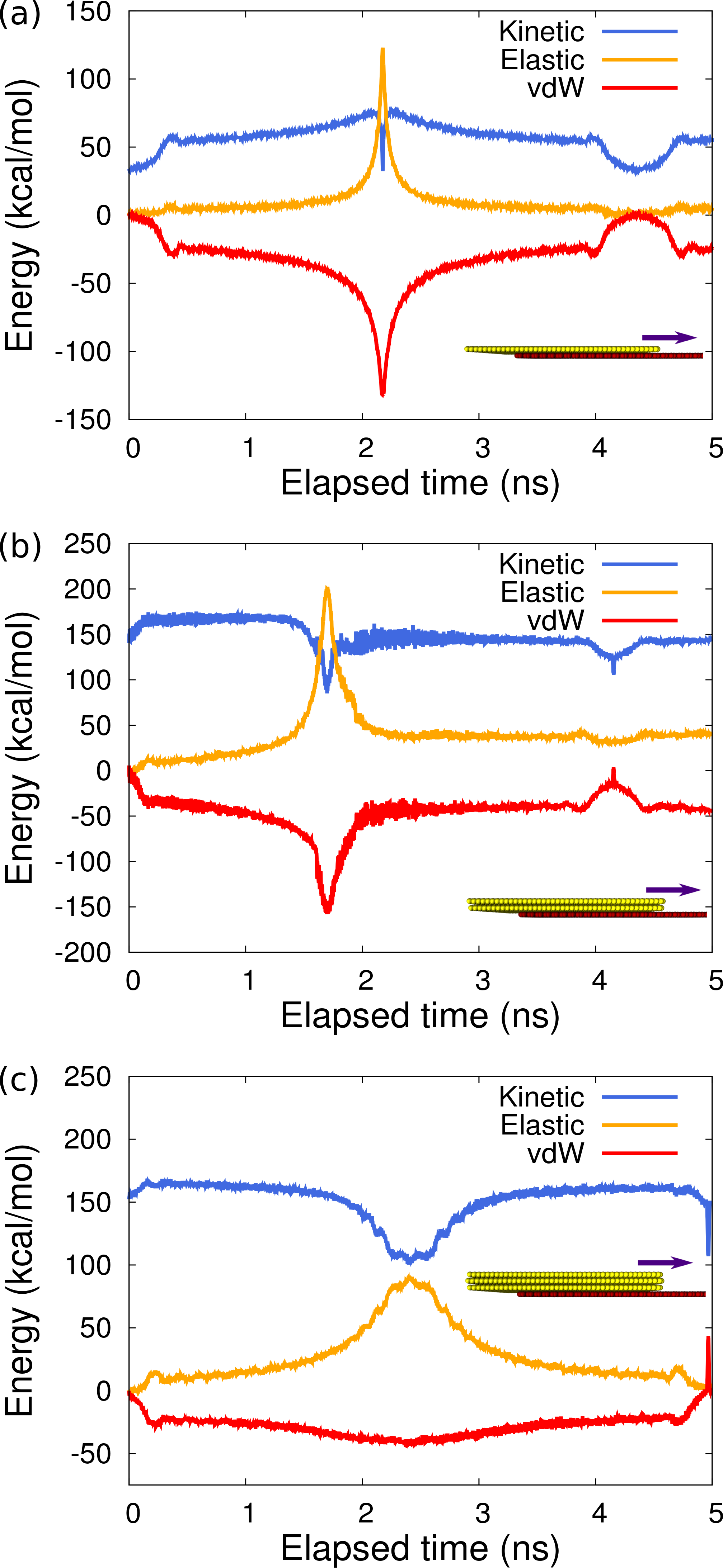}
\caption{(Color online) Evolution of the kinetic, elastic and vdW energies during the operation of a single layer (a), double layer (b), and triple layer (c) GNR oscillator. Comparing these results, we notice that in (c) conversion to elastic energy is more gradual. VdW and elastic energy values refer to differences from the initial amounts.}
\label{fig4}
\end{figure}

The time evolution of the kinetic, elastic, and vdW energies during the operation of a GNR oscillator with one, two, and three zigzag ribbon layers is presented in figure \ref{fig4}. In order to understand the role of the number of layers, let us start by analyzing the TL case (figure \ref{fig4} (c)). Motion is simplest in this case: an initial vdW potential energy is first converted to kinetic and then to elastic potential energy. After 2.4 ns motion is reversed, and the elastic potential energy is converted first to kinetic and then to vdW potential energy. 

Let us now analyze the SL results (figure \ref{fig4} (a)). There are two main differences between this and the TL case: (1) conversion from kinetic to elastic energy is rather small until near the reversing point (at t$= 2.2$ ns); (2) vdW energy becomes more negative as the moving ribbon approaches the spiral's end. To understand (1), notice that graphene monolayers can be bent without extension or compression of $\sigma$ bonds \cite{zhang2011bending}. Because of this, the elastic energy to bend SL graphene is rather low, and it takes rather large curvature to increase its elastic energy. In regards to effect (2), it is actually induced by the large curvature: at the center of the spiral, some atoms in the moving graphene sheet are close enough to mutually attract via vdW interactions. Meanwhile, the local curvature at the reversing point of the trilayer movement is not enough to provide an increase in attractive energy. Effects (1) and (2) also take place for the DL case, figure \ref{fig4} (b). The evolution of the DL and TL oscillators is compared in movie S3 \cite{SM}. 

\begin{figure}
\centering
\includegraphics[width= 3.0 in]{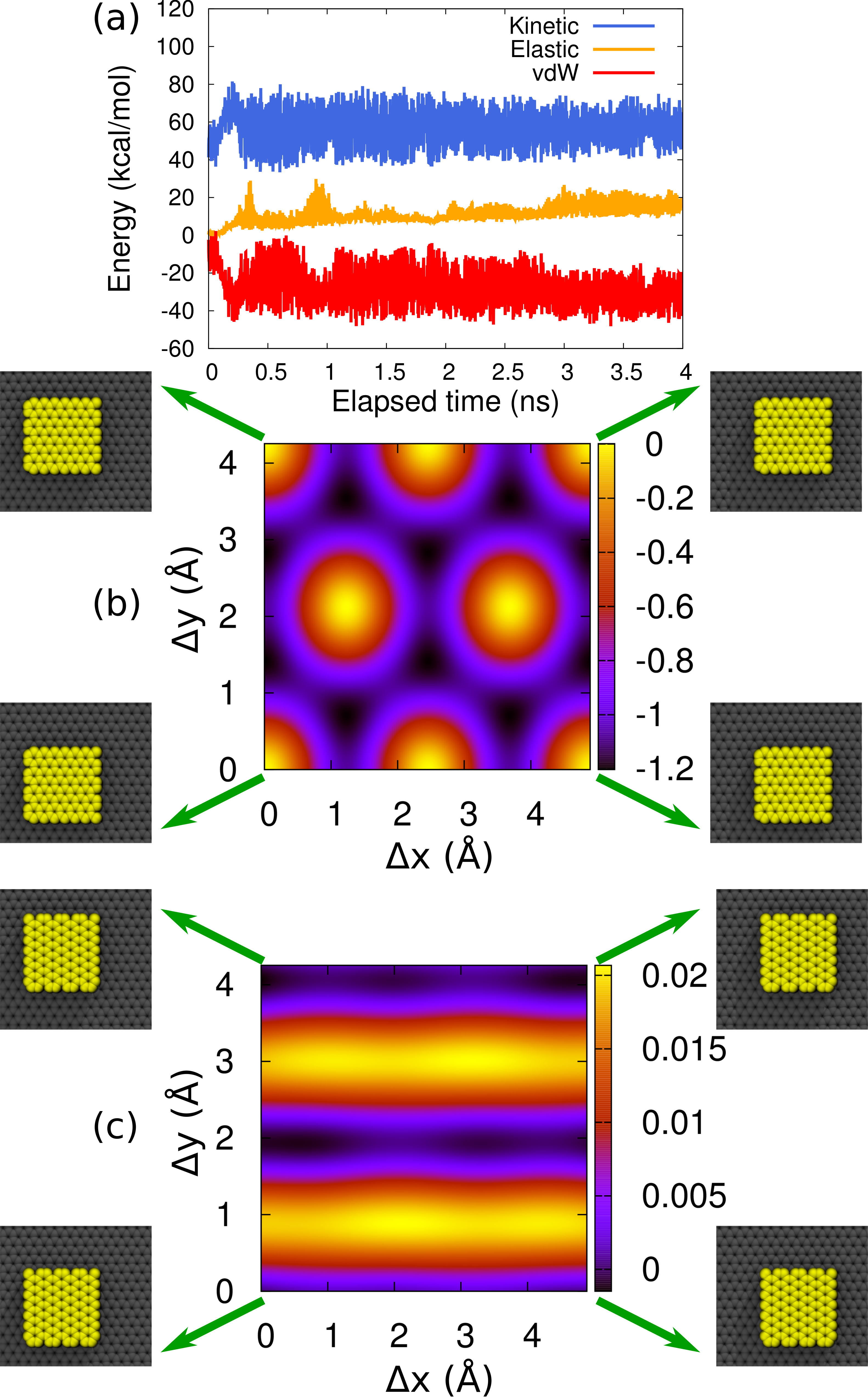}
\caption{(Color online) (a) Time evolution of the kinetic, elastic, and vdW energy terms for a single layer armchair GNR moving on an armchair GNR substrate. Notice the abrupt back and forth conversion between kinetic and vdW potential energies. (b-c) Potential energy maps obtained by placing a graphene flake in 22500 different configurations above an armchair graphene sheet, and measuring the potential energy of each configuration. The flake was armchair in (b) and zigzag in (c). The insets show some of the tested configurations. Energy values are given in kcal/mol. Notice that the energy changes are two orders of magnitude larger in the commensurate case. }
\label{fig5}
\end{figure}

In the GNR simulations discussed so far, all moving ribbons were zigzag, but we also carried out simulations using moving armchair ribbons. The main difference between the two instances is that AB stacking can occur in the latter case, since the spiral substrate is also armchair. When both the substrate and the moving GNR are armchair, the stacking is commensurate; when the moving GNR is zigzag, the stacking is incommensurate. Movie S4 \cite{SM} presents a comparison between dynamics with commensurate and incommensurate stacking. The major difference observed is that there is no oscillation in the commensurate case. The energy evolution during the attempt to drive a commensurate oscillator is presented in figure \ref{fig5}(a). The very beginning of the simulation is similar to the incommensurate case, with conversion of vdW potential energy to kinetic energy. After some time is elapsed, however, the behavior changes, and energy starts to rapidly shift back and forth between vdW potential and kinetic energy. We associate the high kinetic energy/low vdW energy instances with configurations in which AB stacking occurs, and the low kinetic energy/high vdW potential energy instances with configurations in which AA stacking occurs.

The reason why motion is so different is that friction is much higher in the commensurate case \cite{guo2007modifying}. For an incommensurate stacking of graphene layers, in fact, friction almost vanishes \cite{zhang2013superlubricity}, in a phenomenon known as superlubricity \cite{xu2013interaction}. To understand this phenomenon, consider the potential energy maps presented in Figure \ref{fig5} (b) and (c). The method used to obtain these maps was to place a small graphene flake in different positions above a periodic graphene sheet, and then calculating the resulting potential energy. 22500 configurations were used for each map, and every energy calculation corresponds to a dot in this graph. Further details can be found in the supporting information \cite{SM}. Comparing the two maps, we see that energy remains almost constant in the incommensurate graph (Figure \ref{fig5}c). Since forces require gradients of energy to arise, motion across such a near equipotential should be almost frictionless. We also tested smaller graphene flakes, and found that the energy changes decreased for the commensurate case but not for the incommensurate case - see the supporting information \cite{SM}. Conversely, this suggests that the contrast between the two cases would only increase if we tested larger flake sizes. Finally, note that in the dynamical tests friction was very high in the commensurate case by design, as we used a substrate and an oscillator with the same width. Previous MD simulations \cite{liu2014high} showed that a narrow GNR flake moving on a wide graphene sheet rotates to avoid the high friction commensurate configuration. 

That friction can essentially vanish for GNR and CNT oscillators is important, because the curvature of existing spiral nanostructures \cite{blees2015graphene,pevzner2011confinement} is rather low when compared to the ones we used in the simulations. This means that driving forces could also be rather low if such structures were used as templates for the synthesis of the proposed nanoscillators. In such a case, high friction would likely destroy directed motion.

\section{Conclusion}

In summary, our MD simulations show that gradients of surface and bending energies can drive motion. In particular, we demonstrate that it's possible to use CNT/GNR spiral-shaped substrates to drive nanoscillators. We also show how current instances of nanoscale motion driven by system set-ups can be interpreted in terms of gradients of a simple energy expression. Additionally, the former analysis suggests that torsional elastic energy could likewise be used to drive movement. By comparing the motion of commensurate and incommensurate GNR oscillators, we show that proper interlayer orientation is critical to ensure sustained directed movement. The vanishing friction force in the incommensurate case indicates that bending energy driven oscillators could operate with curvature gradients milder than the ones used in our simulations. Such gradients are present in existing nanostructures, suggesting that our proposed \textit{nanomechanotaxis} might be feasible with currently available technology.

\section{Acknowledgements}
We would like to thank the Brazilian agencies CNPq and FAPESP (Grants 2013/08293-7, 2016/18499-0, and 2019/07157-9) for financial support. Computational and financial support from the Center for Computational Engineering and Sciences at Unicamp through the FAPESP/CEPID Grant No. 2013/08293-7 is also acknowledged. This work was financed in part by CAPES - Finance Code 001. LDM acknowledges the support of the High Performance Computing Center at UFRN (NPAD/UFRN).

\bibliography{mybib}

\begin{thebibliography}{10}
\expandafter\ifx\csname url\endcsname\relax
  \def\url#1{\texttt{#1}}\fi
\expandafter\ifx\csname urlprefix\endcsname\relax\def\urlprefix{URL }\fi
\expandafter\ifx\csname href\endcsname\relax
  \def\href#1#2{#2} \def\path#1{#1}\fi

\bibitem{regan2004carbon}
B.~Regan, S.~Aloni, R.~Ritchie, U.~Dahmen, A.~Zettl, Carbon nanotubes as
  nanoscale mass conveyors, Nature 428~(6986) (2004) 924--927 (2004).

\bibitem{zhao2010mass}
J.~Zhao, J.-Q. Huang, F.~Wei, J.~Zhu, Mass transportation mechanism in
  electric-biased carbon nanotubes, Nano letters 10~(11) (2010) 4309--4315
  (2010).

\bibitem{junno1995controlled}
T.~Junno, K.~Deppert, L.~Montelius, L.~Samuelson, Controlled manipulation of
  nanoparticles with an atomic force microscope, Applied Physics Letters
  66~(26) (1995) 3627--3629 (1995).

\bibitem{kudernac2011electrically}
T.~Kudernac, N.~Ruangsupapichat, M.~Parschau, B.~Maci{\'a}, N.~Katsonis, S.~R.
  Harutyunyan, K.-H. Ernst, B.~L. Feringa, Electrically driven directional
  motion of a four-wheeled molecule on a metal surface, Nature 479~(7372)
  (2011) 208--211 (2011).

\bibitem{wang2014motion}
C.~Wang, S.~Chen, Motion driven by strain gradient fields., Scientific reports
  5 (2015) 13675--13675 (2015).

\bibitem{chen2018nanoscale}
P.~Chen, S.~Lv, Y.~Li, J.~Peng, C.~Wu, Y.~Yang, A nanoscale rolling actuator
  system driven by strain gradient fields, Computational Materials Science 154
  (2018) 380--392 (2018).

\bibitem{dai2016nanoscale}
C.~Dai, Z.~Guo, H.~Zhang, T.~Chang, A nanoscale linear-to-linear motion
  converter of graphene, Nanoscale 8~(30) (2016) 14406--14410 (2016).

\bibitem{nourhani2015guiding}
A.~Nourhani, V.~H. Crespi, P.~E. Lammert, Guiding chiral self-propellers in a
  periodic potential, Physical review letters 115~(11) (2015) 118101 (2015).

\bibitem{chang2015nanoscale}
T.~Chang, H.~Zhang, Z.~Guo, X.~Guo, H.~Gao, Nanoscale directional motion
  towards regions of stiffness, Physical Review Letters 114~(1) (2015) 015504
  (2015).

\bibitem{bigoni2015eshelby}
D.~Bigoni, F.~Dal~Corso, F.~Bosi, D.~Misseroni, Eshelby-like forces acting on
  elastic structures: theoretical and experimental proof, Mechanics of
  Materials 80 (2015) 368--374 (2015).

\bibitem{bigoni2014torsional}
D.~Bigoni, F.~Dal~Corso, D.~Misseroni, F.~Bosi, Torsional locomotion,
  Proceedings of the Royal Society A: Mathematical, Physical and Engineering
  Science 470~(2171) (2014) 20140599 (2014).

\bibitem{mackerell1998all}
A.~D. MacKerell, D.~Bashford, M.~Bellott, R.~Dunbrack, J.~D. Evanseck, M.~J.
  Field, S.~Fischer, J.~Gao, H.~Guo, S.~Ha, et~al., All-atom empirical
  potential for molecular modeling and dynamics studies of proteins, The
  Journal of Physical Chemistry B 102~(18) (1998) 3586--3616 (1998).

\bibitem{peng2008measurements}
B.~Peng, M.~Locascio, P.~Zapol, S.~Li, S.~L. Mielke, G.~C. Schatz, H.~D.
  Espinosa, Measurements of near-ultimate strength for multiwalled carbon
  nanotubes and irradiation-induced crosslinking improvements, Nature
  nanotechnology 3~(10) (2008) 626--631 (2008).

\bibitem{zhao2009size}
H.~Zhao, K.~Min, N.~Aluru, Size and chirality dependent elastic properties of
  graphene nanoribbons under uniaxial tension, Nano letters 9~(8) (2009)
  3012--3015 (2009).

\bibitem{machado2013dynamics}
L.~Machado, S.~Legoas, J.~Soares, N.~Shadmi, A.~Jorio, E.~Joselevich,
  D.~Galv{\~a}o, Dynamics of the formation of carbon nanotube serpentines,
  Physical review letters 110~(10) (2013) 105502 (2013).

\bibitem{shadmi2015defect}
N.~Shadmi, A.~Kremen, Y.~Frenkel, Z.~J. Lapin, L.~D. Machado, S.~B. Legoas,
  O.~Bitton, K.~Rechav, R.~Popovitz-Biro, D.~S. Galv{\~a}o, et~al., Defect-free
  carbon nanotube coils, Nano letters 16~(4) (2015) 2152--2158 (2015).

\bibitem{plimpton1995fast}
S.~Plimpton, Fast parallel algorithms for short-range molecular dynamics,
  Journal of computational physics 117~(1) (1995) 1--19 (1995).

\bibitem{SM}
S.~supplementary material at \textit{link} for~simulation movies, further
  details.

\bibitem{koenig2011ultrastrong}
S.~P. Koenig, N.~G. Boddeti, M.~L. Dunn, J.~S. Bunch, Ultrastrong adhesion of
  graphene membranes, Nature nanotechnology 6~(9) (2011) 543--546 (2011).

\bibitem{jawed2014coiling}
M.~K. Jawed, F.~Da, J.~Joo, E.~Grinspun, P.~M. Reis, Coiling of elastic rods on
  rigid substrates, Proceedings of the National Academy of Sciences 111~(41)
  (2014) 14663--14668 (2014).

\bibitem{legoas2003molecular}
S.~Legoas, V.~Coluci, S.~Braga, P.~Coura, S.~Dantas, D.~Galvao,
  Molecular-dynamics simulations of carbon nanotubes as gigahertz oscillators,
  Physical review letters 90~(5) (2003) 055504 (2003).

\bibitem{zhang2011bending}
D.-B. Zhang, E.~Akatyeva, T.~Dumitric{\u{a}}, Bending ultrathin graphene at the
  margins of continuum mechanics, Physical review letters 106~(25) (2011)
  255503 (2011).

\bibitem{guo2007modifying}
Y.~Guo, W.~Guo, C.~Chen, Modifying atomic-scale friction between two graphene
  sheets: A molecular-force-field study, Physical Review B 76~(15) (2007)
  155429 (2007).

\bibitem{zhang2013superlubricity}
R.~Zhang, Z.~Ning, Y.~Zhang, Q.~Zheng, Q.~Chen, H.~Xie, Q.~Zhang, W.~Qian,
  F.~Wei, Superlubricity in centimetres-long double-walled carbon nanotubes
  under ambient conditions, Nature nanotechnology 8~(12) (2013) 912--916
  (2013).

\bibitem{xu2013interaction}
Z.~Xu, X.~Li, B.~I. Yakobson, F.~Ding, Interaction between graphene layers and
  the mechanisms of graphite's superlubricity and self-retraction, Nanoscale
  5~(15) (2013) 6736--6741 (2013).

\bibitem{liu2014high}
Y.~Liu, F.~Grey, Q.~Zheng, The high-speed sliding friction of graphene and
  novel routes to persistent superlubricity, Scientific reports 4 (2014).

\bibitem{blees2015graphene}
M.~K. Blees, A.~W. Barnard, P.~A. Rose, S.~P. Roberts, K.~L. McGill, P.~Y.
  Huang, A.~R. Ruyack, J.~W. Kevek, B.~Kobrin, D.~A. Muller, et~al., Graphene
  kirigami, Nature (2015).

\bibitem{pevzner2011confinement}
A.~Pevzner, Y.~Engel, R.~Elnathan, A.~Tsukernik, Z.~Barkay, F.~Patolsky,
  Confinement-guided shaping of semiconductor nanowires and
  nanoribbons:“writing with nanowires”, Nano letters 12~(1) (2011) 7--12
  (2011).

\end{thebibliography}

\end{document}